\documentclass[prl,aps]{revtex4}

\usepackage{enumerate}
\usepackage{enumitem}
\usepackage{graphicx}
\usepackage{dcolumn}
\usepackage{bm}
\usepackage{graphicx}
\usepackage{amssymb}
\usepackage{epstopdf}

\usepackage[english]{babel}
\newcommand{\LTO}{$\mbox{LaTiO}_3${$\mbox{ } $}}
\newcommand{\STO}{$\mbox{SrTiO}_3${$\mbox{ }  $}}
\newcommand{\LAO}{$\mbox{LaAlO}_3${$\mbox{ }  $}}
\newcommand{\TiO}{$\mbox{TiO}_2${$\mbox{ }  $}}
\newcommand{\LO}{$\mbox{LaO}${$\mbox{ }  $}}

\begin{document}

\title{Two-dimensional superconductivity induced by high-mobility carrier doping in \LTO/\STO hetero-structures}

\author{J. Biscaras$^1$, N. Bergeal$^1$, S. Hurand$^1$, C. Grosset\^ete$^1$, A. Rastogi$^2$, R. C. Budhani$^{2,3}$, D. LeBoeuf$^4$, C. Proust$^4$, J. Lesueur$^1$}

\affiliation{$^1$LPEM- UMR8213/CNRS - ESPCI ParisTech - UPMC, 10 rue Vauquelin - 75005 Paris, France}

\affiliation{$^2$Condensed Matter - Low Dimensional Systems Laboratory, Department of Physics, Indian Institute of Technology Kanpur, Kanpur 208016, India}
\affiliation{$^3$National Physical Laboratory, New Delhi - 110012, India }
\affiliation{$^4$Laboratoire National des Champs Magn\'etiques Intenses, UPR 3228, (CNRS-INSA-UJF-UPS), Toulouse 31400, France }

\date{\today}

\begin{abstract}
In this letter, we show that a superconducting two-dimensional electron gas is formed at the  \LTO/\STO interface whose transition temperature can be modulated by a back-gate voltage. The gas consists of two types of carriers : a majority of low-mobility carriers always present, and a few high-mobility ones that can be injected by electrostatic doping. The calculation of the electrons spatial distribution in the confinement potential  shows that the high-mobility electrons responsible for superconductivity set at the edge of the gas whose extension can be tuned by field effect.  
\end{abstract}

\maketitle

Oxide based heterostructures appear as serious challengers for future electronics as they offer a great variety of electronic orders suitable to achieve new functionalities\cite{Takagi:2010p9802,Mannhart:2010p6675}. A key issue for the oxide electronics to emerge is the ability of tuning these properties with an electric field. The discovery of high mobility two-dimensional electron gas (2DEG) in oxide heterostructures is a milestone on this road\cite{Ohtomo:2004p442}. \STO based structures attracted much attention in this context, since \emph{(i)} large mobility can be obtained,  \emph{(ii)}  their complex phase diagram includes various electronic orders such as superconductivity\cite{Reyren:2007p214,Biscaras:2010p7764,Kozuka:2009p2302} and magnetism\cite{brinkman:2007p493,bert:2011p767,li:2011p762} and \emph{(iii)}  their carrier density can be electrostatically modulated \cite{Thiel:2006p2271,Caviglia:2008p116}. However, the electrostatic control of the 2DEG properties is not fully understood yet.
Here we show that superconductivity at \LTO/\STO  interface can be turned on and controlled by injection of a few highly mobile electrons.\\
\indent The \LTO/\STO perovskite heterostructure is particularly interesting in the context of \STO based interfaces since it is made of  \TiO planes as building blocks, where the Ti atoms can have multiple valence values. 
\STO is a well known band insulator of 3.2~eV bandgap with Ti atoms in the $3d^0$ (4$^+$) configuration. On the other hand, \LTO is an antiferromagnetic  Mott insulator, with Ti atoms in the $3d^1$ (3$^+$) configuration. Therefore, when a thin layer of  \LTO is epitaxially grown on a  \TiO terminated \STO  substrate,  half an electron per unit cell is available to form a 2DEG confined at the interface, as confirmed by band structure calculations\cite{Larson:2008p963}. We have shown recently that the electron gas extends a few unit cells  in the \STO layer and undergoes a superconducting transition at low temperature ($\sim$ 200~mK)\cite{Biscaras:2010p7764}. In this letter, we investigate the phase diagram of \LTO/\STO interfaces by electronic transport measurements at low temperature (20~mK) and high field (45~T) as a function of a back-gate voltage. \LTO epitaxial layers were grown on (100) \STO single crystals using Pulsed Laser Deposition as described in reference \cite{Biscaras:2010p7764}. Samples are cut in rectangular shapes along the two orthogonal  in plane directions, hereafter referred as xx and yy. A metallic titanium layer was deposited at the rear of the 0.5~mm thick \STO substrate, to form an electrostatic back-gate. After the sample is cooled down, the gate voltage $V_\mathrm{G}$ is ramped up to +200~V. This procedure insures that all voltage sweeps made subsequently are reversible and reproducible.

\begin{figure}[h]
\includegraphics[width=12cm]{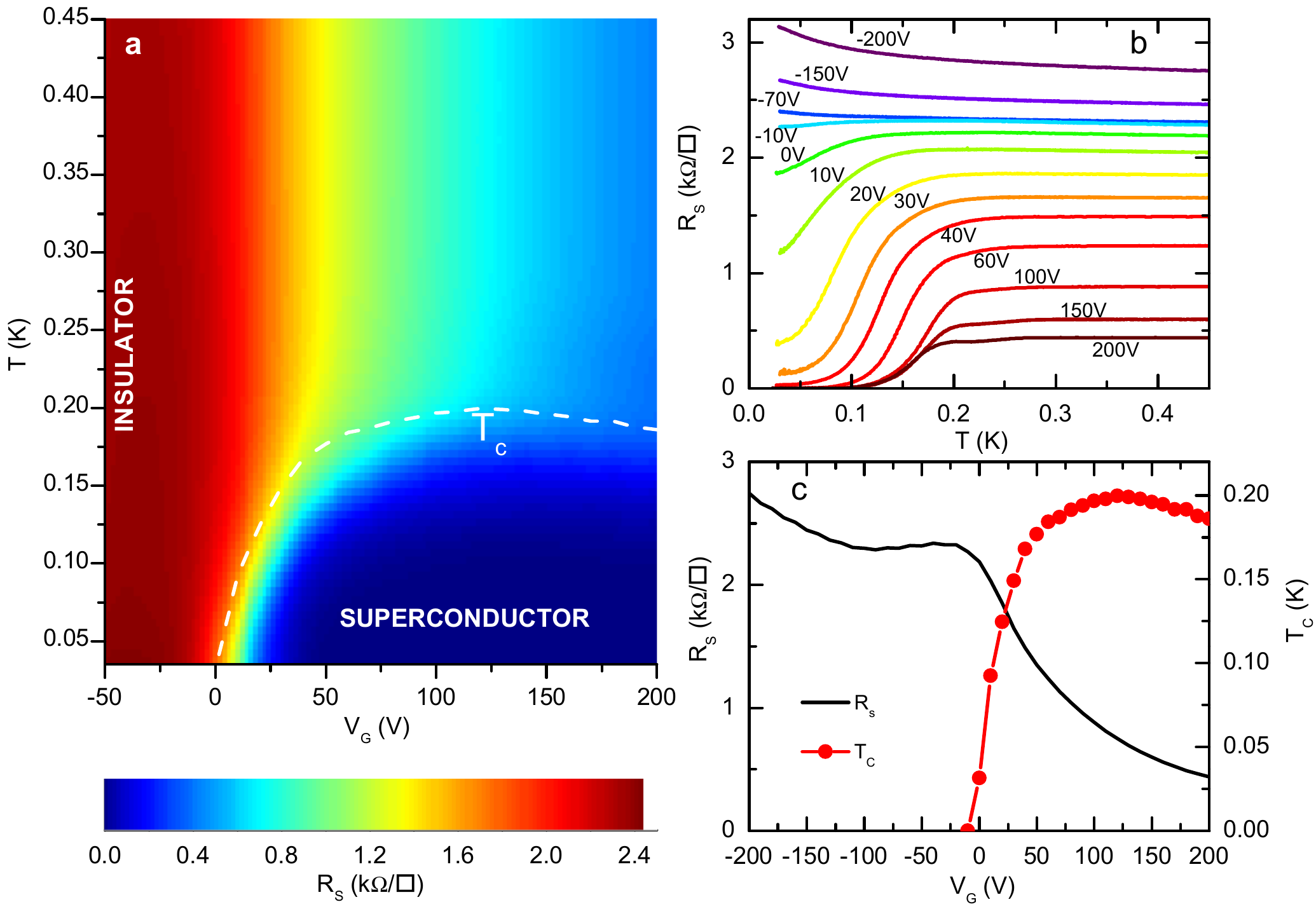}
    \caption{Superconducting transition as a function of gate voltage for sample A (XX direction). {\bf a}, Sheet resistance of sample A  in color scale as a function of temperature and gate voltage from  -50~V to +200~V, the superconducting transition temperature (defined as a 15\% drop in sheet resistance) is superimposed in white dashed line (see Supplementary Information Part I for a discussion on $T_c$ criterion). {\bf b}, Sheet resistance as a function of temperature for selected gate voltages. {\bf c}, Sheet resistance at 450~mK (left scale) and superconducting transition temperature (right scale) as a function of gate voltage (see Supplementary Information Part II for the YY direction).}
\label{fig1}
 \end{figure}

Fig.1a and 1b display the sheet resistance $R_\mathrm{S}$ of a 15 unit cells thick sample (sample A) as a function of temperature for different gate voltages $V_{\mathrm{G}}$. The normal state resistance, its temperature dependence and the superconducting critical temperature $T_c$ strongly depend on $V_{\mathrm{G}}$. A superconducting to insulator transition  takes place as carriers are removed from the 2DEG. $T_c$ has a maximum of $\sim$200~mK when adding electrons, and is suppressed when the gate voltage becomes negative.  This suppression is correlated to a singular behaviour of the normal resistance near $V_{\mathrm{G}}$=0 (Fig.1c), indicating that the gate voltage not only controls the carrier density, but also modifies deeply the electronic properties of the 2DEG.\\

The Hall resistance $R_\mathrm{Hall}$ was measured at low magnetic field ($B<$5~T) as a function of gate voltage. The apparent Hall electron density $n_\mathrm{Hall} = B/eR_\mathrm{Hall}$, which ranges from 3 to 5$\times 10^{13}$~$\mbox{cm}^{-2}$, starts rising from $V_{\mathrm{G}}$=-200~V as expected for negative carriers, but drops surprisingly for positive gate voltage (Fig.2b). The mobility $\mu_\mathrm{Hall} = 1/e n_\mathrm{Hall} R_\mathrm{S}$ is constant for $V_{\mathrm{G}}<0$, and then rises abruptly for $V_{\mathrm{G}}>0$ (Fig.2c purple squares). Measurements performed at higher magnetic field (45~T) reveal that the Hall resistance is linear only for negative $V_{\mathrm{G}}$, and not for positive $V_{\mathrm{G}}$ (Fig.2a). In the latter case, two distinct slopes are evidenced at low and high field respectively, suggesting a multi-band transport scenario. In Fig. 2a., the Hall resistance at high magnetic field  has been fitted with a two-band model :

\begin{eqnarray} 
R_{\mathrm{Hall}}=\frac{B}{e}\frac{\frac{n_1\mu^2_1}{1+\mu^2_1B^2}+\frac{n_2\mu^2_2}{1+\mu^2_2B^2}}{\Big[\frac{n_1\mu_1}{1+\mu^2_1B^2}+\frac{n_2\mu_2}{1+\mu^2_2B^2}\Big]^2+\Big[\frac{n_1\mu^2_1B}{1+\mu^2_1B^2}+\frac{n_2\mu^2_2B}{1+\mu^2_2B^2}\Big]^2}
\label{eqn1}
\end{eqnarray}  

where $n_1$ and $n_2$ are the 2D electron densities and, $\mu_1$ and $\mu_2$ the corresponding mobilities, with the constraint $1/R_s= en_1 \mu_1 + e n_2 \mu_2$. As reported in Fig.2b and 2c, low and constant mobility $\mu_1$ carriers (hereafter refered as LMC) are present for all gate biases, whereas a few highly mobile electrons (hereafter refered as HMC)  with a mobility $\mu_2$ increasing linearly with bias, show up for positive $V_{\mathrm{G}}$ only, as also observed by Kim {\em et al.} \cite{Kim:2010p9791} . As expected for an electrostatic doping, the total number of carriers $n_\mathrm{Total}=n_1+n_2$ rises monotonically with $V_{\mathrm{G}}$ (from 3 to 7$\times 10^{13}$~$ \mbox{cm}^{-2}$).\\

Such gate voltage dependent behaviour has been measured in all our samples, as for example the one reported in Fig.3 (sample B), whose total carrier density is slightly higher (from 6 to 10$\times 10^{13}$~$\mbox{cm}^{-2}$). The dependence of $n_\mathrm{Total}$ with $V_{\mathrm{G}}$ has been confirmed by measuring the capacitance $C(V_{\mathrm{G}})$ between the back-gate and the 2DEG and integrating it over the voltage range to obtain the electrostatic sheet carrier density
\begin{eqnarray} 
n_S(V_{\mathrm{G}})=n_S(V_{\mathrm{G}}=-200 \mathrm{V})+\frac{1}{eA}\int_{-200}^{V_{\mathrm{G}}}C(V)dV
\end{eqnarray}  
where $A$ is the area of the capacitor. As shown in Fig.3a, $n_S$ superimposes perfectly on $n_\mathrm{Hall}$ at negative voltage, as expected since there is only one type of carrier. At positive voltage, it matches $n_\mathrm{Total}$ extracted from the two-band model analysis of high field measurements. The sigmoid shape of the total number of carriers at low temperature is characteristic of the dielectric constant $\epsilon_R$ of the \STO substrate, which is highly non-linear and strongly temperature dependent below $\sim$50~K as \STO undergoes a transition to a quantum paraelectric phase\cite{NEVILLE:1972p3397,HEMBERGER:1995p2310}. At higher temperature, $C(V_{\mathrm{G}})$ is constant, the total number of carriers rises linearly with $V_{\mathrm{G}}$ and equals the Hall number of carriers in the entire voltage range ($n_\mathrm{Total}=n_\mathrm{Hall}$), indicating that no HMC are present in the 2DEG. As seen in  Fig.3b, the mobility is also constant at this temperature, with the same value that the one measured a low temperature and negative $V_{\mathrm{G}}$. When lowering the temperature below 50~K, the low field Hall mobility rises up by more than an order of magnitude at $V_{\mathrm{G}}$=200V, and the low field Hall carrier density departs from the integrated capacitance measurements for $V_{\mathrm{G}}>0$ only. This is a strong indication that the two-bands scenario is intrinsically related to the non-linear variation of \STO dielectric constant $\epsilon_R$ in the quantum paraelectric regime. \\

\begin{figure}[h]
\includegraphics[width=12cm]{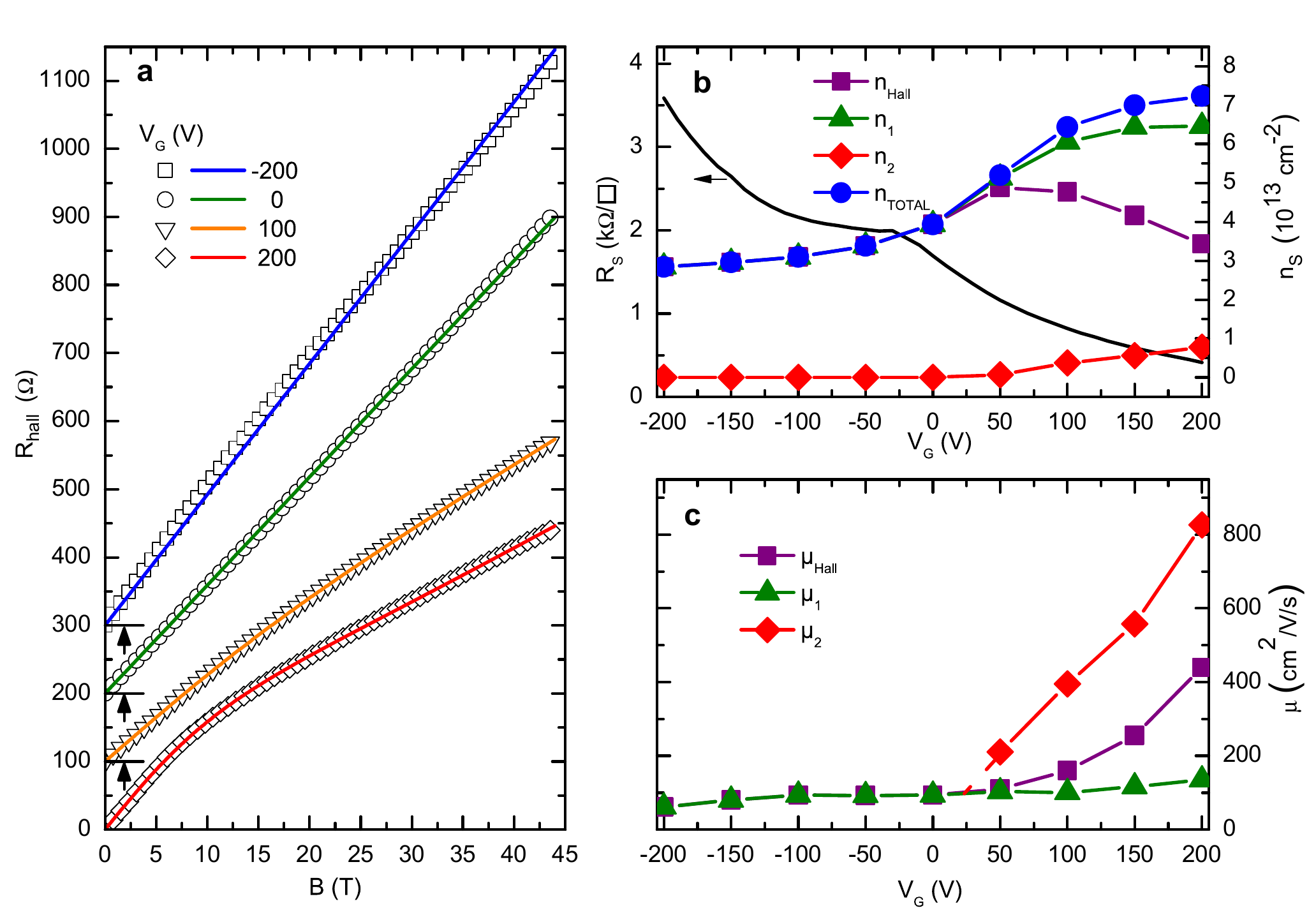}
    \caption{Hall effect and two carriers analysis. {\bf a}, Hall resistance as a function of magnetic field for different gate voltages measured at 4.2~K on sample A (YY direction). Open symbols correspond to experimental data, and full lines to fits with expression (\ref{eqn1}). An offset, indicated by a black horizontal segment and an arrow, has been added to separate the curves. {\bf b}, Sheet resistance (left scale) and carrier density (right scale) extracted from (\ref{eqn1}) as a function of gate voltage measured at 4.2~K. $n_1$ (resp. $n_2$) corresponds to LMC (resp. HMC), $n_\mathrm{Total}$  to the total density of carriers and $n_\mathrm{Hall}$ to the low field Hall number. {\bf c}, Mobilities $\mu_1$, $\mu_2$ and $\mu_\mathrm{Hall}$ corresponding to LMC, HMC and low field Hall number extracted from (\ref{eqn1}).}
    \label{fig2}
    \end{figure}
    
To model the  2DEG properties, we have solved  the coupled Schr\"odinger and Poisson equations self consistently, taking into account the dependence of $\epsilon_R$ with $V_{\mathrm{G}}$ and the continuity of the potential in the whole sample. Indeed, the gate voltage not only controls the total number of carriers in the 2DEG, but also the profile of the conduction band in the substrate. As opposed to previous calculations\cite{Copie:2009p5635,Bell:2009p6086,Ueno:2008p5814}, we do not use a wedge-shape potential to model the confining potential of the 2DEG at the interface, since it cannot provide a continuous solution with the bulk band-bending. Details of the calculations are given in the Supplementary Information Part III. The self consistent potential profile, as depicted for different $V_{\mathrm{G}}$ in Fig.4a to 4c for sample A, is made of a deep potential well of typical width 2~nm, which accommodates a few sub-bands located in the vicinity of the interface (see Supplementary Part V for sample B). For positive gate voltage, higher sub-bands filled up to the top of the conduction band profile, extend up to 7~nm within the \STO substrate. The corresponding carrier density profiles show a dominant contribution of carriers located at the interface, with a pronounced 2D character and therefore more subject to localization, which we assign to the LMC observed experimentally, and a minor contribution of more delocalized carriers at positive $V_{\mathrm{G}}$, corresponding to the HMC. The extension of the gas reported in Fig.4d  is consistent with experimental results obtained on \STO based interfaces  \cite{Biscaras:2010p7764,Reyren:2007p214,Copie:2009p5635,Dubroka:2010p6353}. It increases significantly for $V_{\mathrm{G}}>0$, as does the measured HMC density.\\

The origin of this peculiar behaviour is deeply related to the \STO dielectric constant $\epsilon_R$ variations, which is highly electric field dependent at low temperature, and therefore changes locally within the heterostructure. As reported in Fig.4a,b,c, the potential well is deep at the interface, and the corresponding electric field so high that $\epsilon_R$ decreases to its lower value whatever the gate voltage is, leading to a strong confinement of the gas. For $V_{\mathrm{G}}<0$, the band-bending reinforces this feature. For $V_{\mathrm{G}}>0$, the edge of the well becomes shallower, the local electric field smaller, and $\epsilon_R$ starts rising towards the bulk, enhancing the de-confinement of the carriers. Their mobility gradually increases as the positive voltage is raised  because the 2DEG extends away from the interface where the scattering is stronger \cite{Seo:2009p9442} and anisotropic \cite{noteAniso20} (see suplementary part II). This mechanism does not apply at high temperature when $\epsilon_R$ becomes field independent, and LMC are solely observed above 60~K, as reported here (see Figure~3). According to our model, the gate voltage not only controls the carrier density, but also the 2DEG spatial extension and its transport properties. The vicinity of $V_\mathrm{G}$=0 appears to be a turning point beyond which, for positive bias, mobile carriers are injected away from the interface. As can be seen in Fig.4d, superconductivity is intimately related to the appearance of HMC and their de-confinement within the \STO substrate.\\

\begin{figure}[h]
\includegraphics[width=12cm]{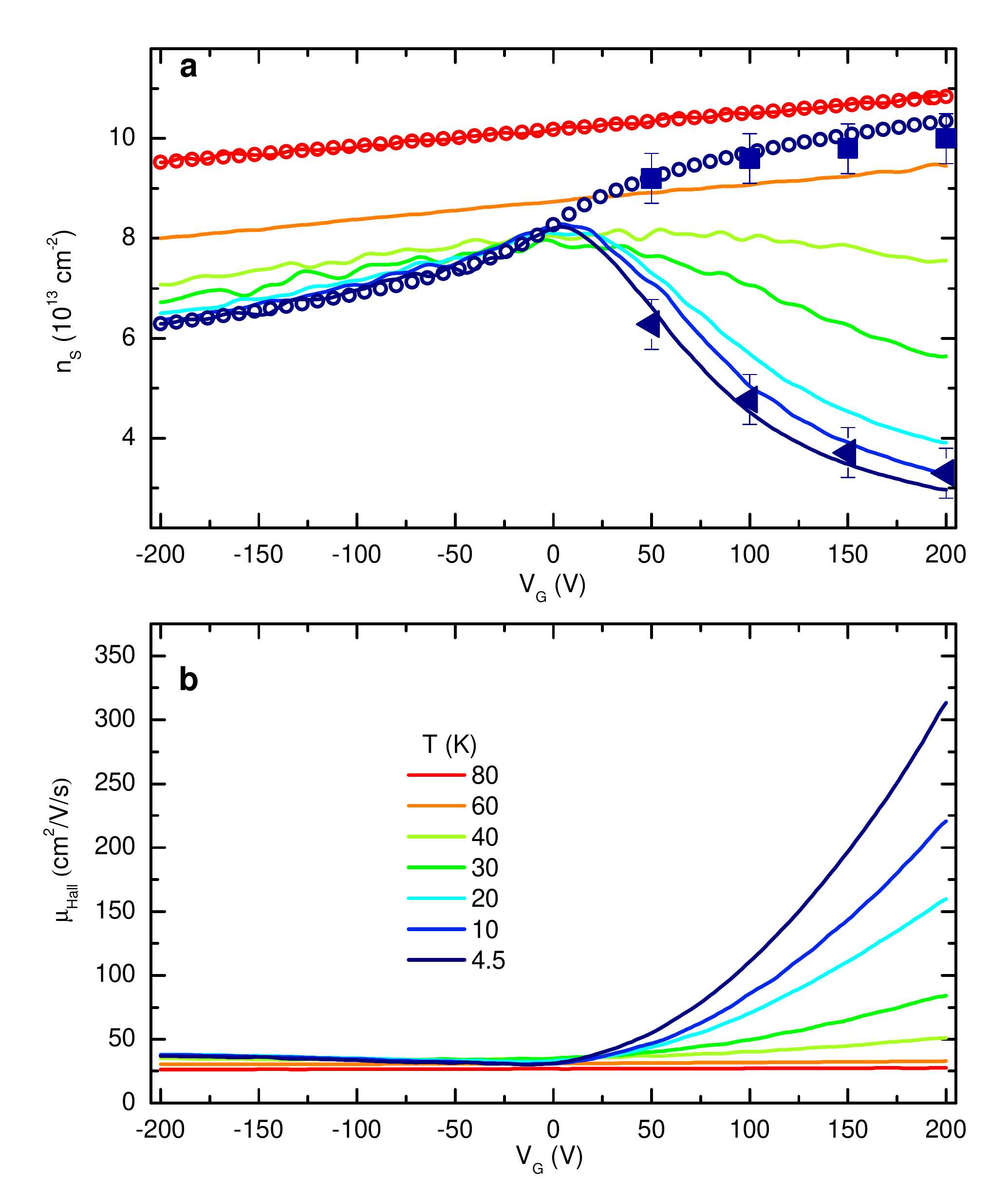}
    \caption{Evolution of carrier density and mobility as a function of gate voltage and temperature. {\bf a}, Low field (5~T) Hall number of carriers  (full lines) and total number of carriers extracted from capacitance measurements (open symbols) as a function of gate voltage measured at different temperatures on sample B. Total number (squares) and low field Hall number (triangles) extracted from the two carriers analysis of high field experiments (45~T) are superimposed to the other curves, showing a very good agreement. {\bf b}, Hall mobility corresponding to low field measurements as a function of gate voltage at the same temperatures. Beyond 60~K, that is when the non-linearity of $\epsilon_R$ strongly decreases, $n_S$ and $\mu_{\mathrm{Hall}}$ recover a regular behavior with the gate voltage, that is a rather constant mobility and a linear increase of the carriers concentration expected for a single carrier fluid. }
    \label{fig3}
    \end{figure}

	At first sight, the dome shape of $T_c$ as a function of gate voltage observed at the \LTO/\STO interface may look like the bulk doped \STO one \cite{Koonce:1967p1560}. However, the 3D carriers concentration at the interface (Fig.4a,b,c), even for the most negative gate bias, always reaches the range where bulk superconductivity is observed (i.e. $10^{19}$ to 5$\times 10^{20}$~$\mbox{cm}^{-3}$), whereas our samples are superconducting only when HMC are present. There is clearly something specific to the interface and to the carrier distribution profile. We can rule out a dimensionality driven transition, since the Fermi wavelength of the 2DEG is always larger than the gas expansion according to our model : the systems remains 2D in the whole gate voltage range explored here. On the other hand, we see in Figure 4.c, that as the gas extends within the substrate, i.e. at positive $V_{\mathrm{G}}$, $\epsilon_R$ approaches the bulk \STO value in the region where HMC set. Therefore in this region, the situation could be similar to the one of  bulk doped \STO, with a finite carrier concentration in \STO of high dielectric constant. The mechanism of superconductivity in \STO is still under debate, but most of the theoretical models \cite{Appel:1966p9862,TAKADA:1980p9911,Koonce:1969p1777} take into account the polar properties of \STO and its correlated peculiar dielectric constant to explain the superconducting properties and the dome-shaped phase diagram. Along these lines, superconductivity would appear at positive gate voltage in our \LTO/\STO hetero-structures because the gas extends rather deep in the \STO substrate, and $T_c$ would decrease at high voltage as it does in bulk material. In that scenario, a full microscopic calculation of $T_c$ with varying $\epsilon_R$ is needed, which is not available yet to compare with. Our data also point toward another key ingredient in the problem, the strong scattering at the interface. It is well known that disorder suppresses superconductivity through the enhancement of localization and electron-electron interactions (see for instance \cite{Finkelshtein:1987p5892}). This may explain why for negative gate voltages, when only LMC are present at the interface, the 2DEG no longer display superconductivity. On the other hand, superconductivity is rapidly restored as soon as HMC are injected. Even if this situation looks like a regular disorder-driven superconductor to insulator transition \cite{Goldman:2010p2440}, the presence of two fluids spatially separated as depicted here is a unique situation, which has not yet been described theoretically.  \\

	Multiple carriers behavior has been reported in \LTO/\STO hetero-structures \cite{Kim:2010p9791,Ohtsuka:2010p9619}. The complex band structure of doped bulk \STO \cite{MATTHEISLF:1972p8588} is a natural source of multi-carriers transport properties. Moreover, electronic reconstruction calculations at \LTO/\STO interfaces clearly show that the band structure evolves within a few unit cells from the interface : the weight of Ti $3d_{xy}$ orbitals with a strong 2D character decays over 1 or 2 unit cells \cite{Larson:2008p963,Okamoto:2004p4286,Popovic:2005p9417}. In this case, LMC and HMC would refer to carriers belonging to different sub-bands spatially distinct on a very short scale, less than a nanometer, which is 5 to 10 times shorter than our estimation for the HMC de-confinement. Other \STO based hetero-structures also display multi-carriers transport properties \cite{Bell:2009p6086,Seo:2009p9442,Hotta:2007p9923} and de-confined HMC on the 5-10 nm scale \cite{Dubroka:2010p6353}, whereas calculations also point toward short scale (1 to 2 unit cell) electronic reconstruction \cite{Popovic:2008p678}. Therefore, the occurrence of deconfined HMC, whose properties do not strongly depend on the details of the surface reconstruction, is a rather general feature of \STO based hetero-structures or even \STO surface \cite{Meevasana:2011p9019} and the two types of carriers are not simply related to bulk doped \STO multibands. Moreover, when electrostatic doping is used, HMC appear at positive gate voltage in most systems \cite{Bell:2009p6086,Kim:2010p9791}, in line with the model we present in this article. Finally, besides the present work, superconductivity as a function of gate voltage has been reported in \LAO/\STO hetero-structures \cite{Caviglia:2008p116,Bell:2009p6086}, with uneven results, since the superconducting dome closes at different gate voltages in those experiments, and in electric double layer gated with organic electrolyte \STO surfaces\cite{Ueno:2008p5814}, where $T_c$ surprisingly does not depends on the gate voltage. It is therefore difficult to make a direct comparison with our data. Let us emphasize that the behavior we have described here, that is the occurrence of superconductivity when HMC appear is robust, and has been observed in different \LTO/\STO samples. \\

\begin{figure}[h]
\includegraphics[width=12cm]{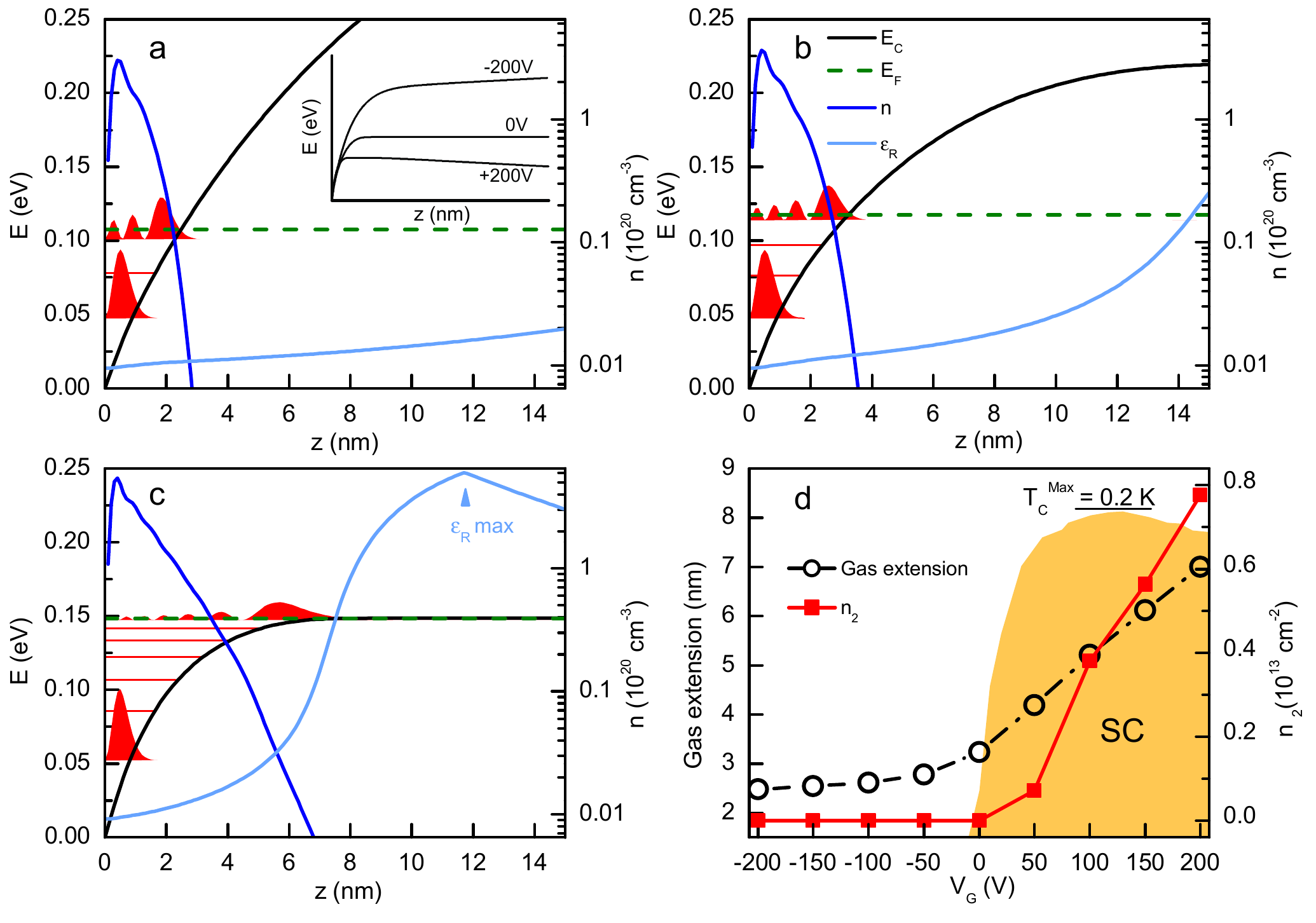}
    \caption{Sub-bands filling calculation in the interface potential well. {\bf a}, {\bf b} and {\bf c}, \STO band bending calculation for three different gate voltages (-200~V, 0~V, +200~V) obtained with the experimental sheet carrier densities of Fig.2b. The figure displays the conduction band profile $E_{\mathrm{C}}$ (black), the Fermi energy $E_{\mathrm{F}}$ (green dashed), the sub-bands energies (red) and the carrier density $n$ (deep blue, left scale) as a function of depth $z$ from the interface. The dielectric constant $\epsilon_R$ as a function of $z$  is drawn on a log scale from 260 to 26000 (see Supplementary figure 4 for precise scale). The square modulus of the envelope function of the first and last filled sub-bands are indicated in arbitrary units (red areas). Inset : conduction band profile $E_{\mathrm{C}}$ at larger scale (0 - 100~nm) for the three different gate voltages. {\bf d}, 2DEG extension from the interface taken as the crossing point of $E_{\mathrm{C}}$ and $E_{\mathrm{F}}$ (left scale) and carrier density $n_2$ of HMC (right scale) as a function of gate voltage. The superconducting transition temperature as measured in Fig.1c, is superimposed with the same scale as a reminder.}
    \label{fig4}
    \end{figure}

In summary, superconductivity in \LTO/\STO interfaces can be modulated with a back-gate voltage. In that case, we show that superconductivity is strongly correlated to the injection of highly mobile carriers at the edge of the 2DEG, whose extension increases rapidly with positive gate voltage, because of the conduction band bending of \STO. This shed a new light on the phase diagram of the 2DEG at \LTO/\STO interface, and more generally on the other \STO based interfaces, since the gate voltage appears not only as a way of tuning the carriers density, but also the geometry of the gas. Clearly, 2DEG transport properties appear as a combination of those of carriers rather ``deep'' (a few nanometers) in bulk \STO, and those of carriers next to the interface, and therefore more sensitive to the details of the surface and/or electronic reconstructions \cite{Seo:2007p9926}. That may explain why, if the overall picture among the different systems looks similar, striking differences can occur, such as the observation of magnetism by transport measurements in \LAO/\STO interfaces [7,9] on the contrary to \LTO/\STO ones. The occurrence of different electronic orders (superconductivity, ferromagnetism, antiferromagnetism ...), their interactions and/or competitions, together with the transition to an insulating state may depend on the carriers density and on the gas extension tuned by the gate voltage. Our results opens the way to a new description of the superconductor to insulator transition in these systems, and beyond, to a comprehensive understanding of the 2DEG physics at oxides interfaces.\\

The Authors gratefully thank L. Benfatto, M. Grilli, S. Caprara and A. Santander-Syro for stimulating discussions. This work has been supported by the R\'egion Ile-de-France in the framework of CNano IdF and Sesame program. Part of this work has been 
supported by Euromagnet II. Research in India was funded by the Department of Information Technology, Government of India.\\


\newpage

\large{\textbf{Supplementary Material}}\\

\normalsize
\textbf{Part I : Superconductivity dome.}\\

	The same resistivity as a function of gate voltage and temperature data as in Figure 1 of main text are displayed with the sheet resistance normalized by its value at 450~mK. Therefore, each color corresponds to a given $R_s/R_n$ ratio. Whatever the chosen ratio to define Tc is, a dome like behavior with $V_\mathrm{G}$ is observed, with a decrease of $T_c$ beyond $V_\mathrm{G} \sim$ 130~V. Moreover, the choice of the $R_s/R_n$ criteria to define $T_c$ has little impact on the determination of the gate voltage where $T_c$ goes to zero : $V_\mathrm{G} \sim$ 0 $\pm$ 15~V. Two distinct criteria (15\% and 50\% resistance drop) are enlightened with dashed white and black lines in the Figure.\\
	\begin{figure}[h!]
\includegraphics[width=10cm]{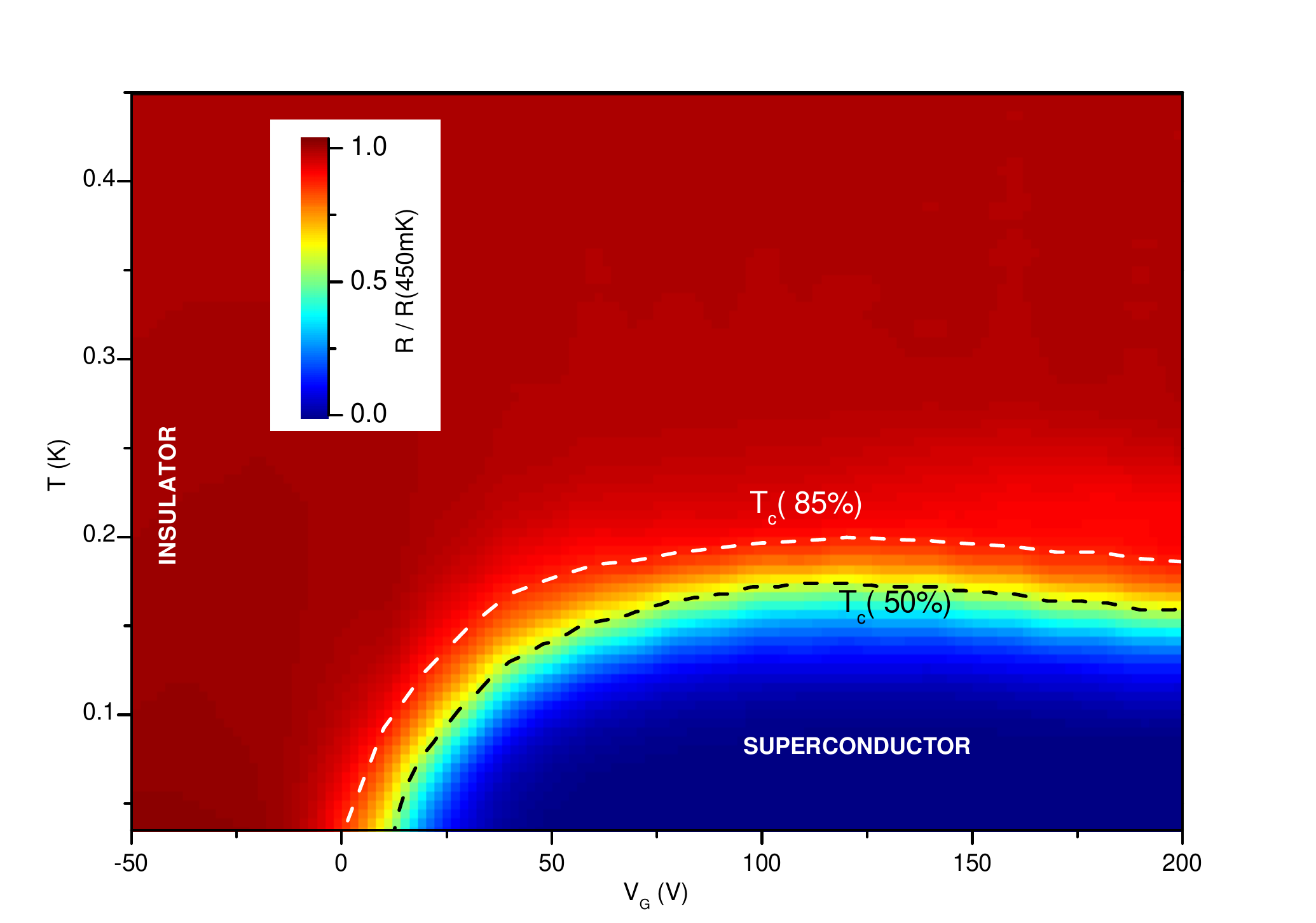}
\caption{Superconducting dome. Normalised sheet resistance as a function of temperature and gate voltage from -50~V to +200~V. The superconducting transition temperatures defined as a 15\% drop in sheet resistance (white dashed line) and 50\% (black dashed line) are superimposed. }
\end{figure}

\textbf{Part II : Anisotropic scattering of  \LTO/\STO samples.}\\

Hetero-structures resistances have been measured along two perpendicular (001) in plane directions XX and YY for samples A and B, as shown in supplementary Figure 2. In all cases, the resistivity displays a kink as a function of gate voltage near $V_\mathrm{G}$ = 0, which appear to be a very robust feature in the \LTO/\STO system.\\
\indent Figure 2  shows that the anisotropy in the resistance is mostly observed at negative gate voltages where the resistance is higher. That means that strong scattering centers are associated with anisotropic scattering. Surface reconstruction is a natural source of anisotropy, which has been extensively studied in \STO (see for instance Deak {\em et al.} [S1]). This is therefore a strong indication that the interface is the place where the scattering is stronger. This is consistent with our scenario where LMC set next to the interface, and are therefore more sensitive to the microscopic details of it. The anisotropy is decreased, or even suppressed at positive gate voltage as HMC are deeper in the \STO and are  less affected by the interface anisotropy. \\
\begin{figure}[h!]
\includegraphics[width=11cm]{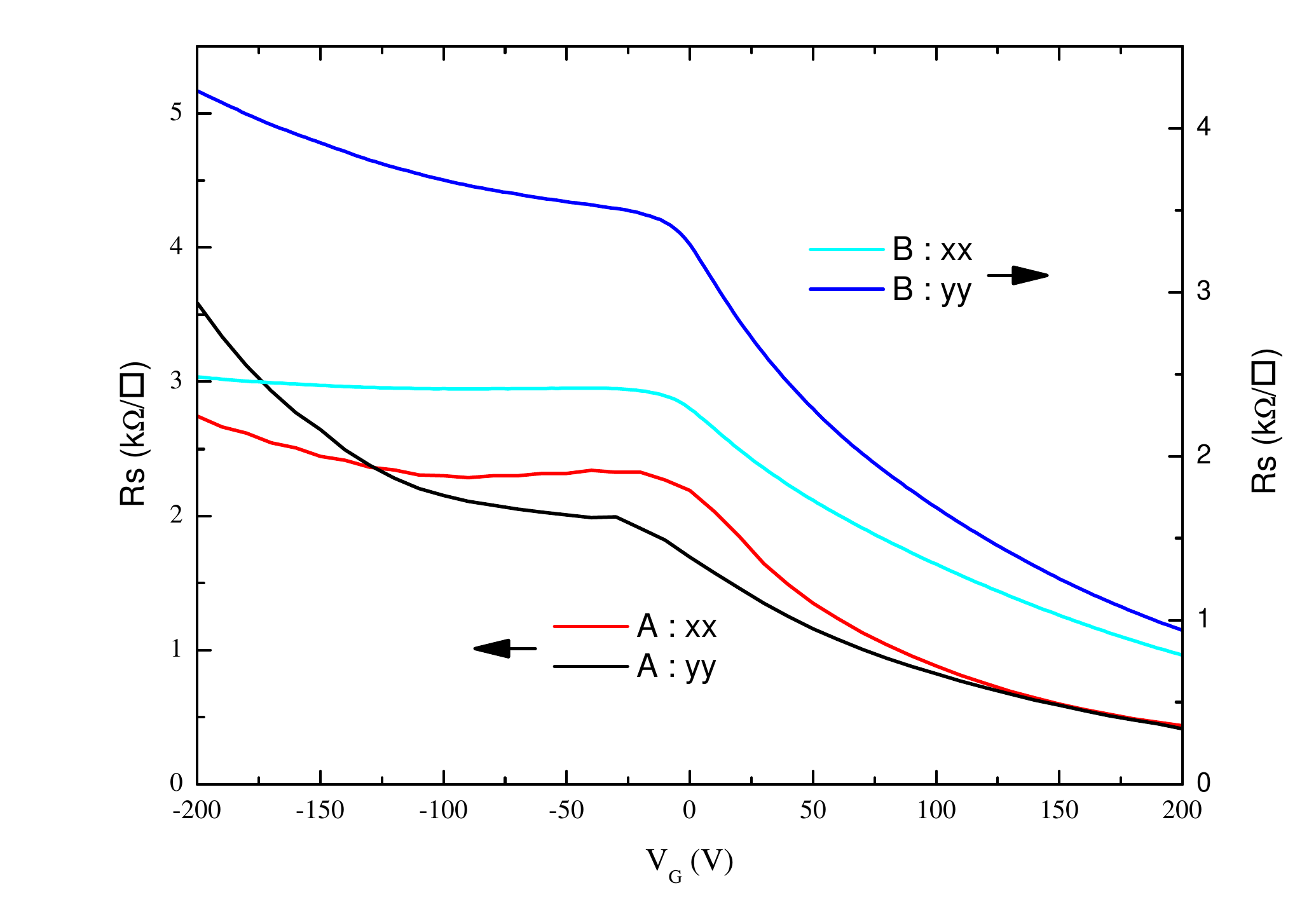}
\caption{Anisotropy of the resistivity as a function of gate voltage. Sheet resistance as a function of gate voltage for two perpendicular directions of sample A (left scale) and sample B (right scale).}
\end{figure}

\indent The reason why is the scattering is more important at the interface is still under debate. As proposed by Seo {\em et al.} [S2] the change in dielectric constant may affect the screening of defects.  However, since the electric field at the interface is huge whatever the gate voltage is, the dielectric constant very close to the interface is basically constant. Therefore, we do not expect a dramatic change in defect screening for LMC when varying the gate voltage. On the contrary, HMC set away from the interface, where the dielectric constant is higher, and therefore the screening of defects better. The deeper the 2DEG expanses within \STO, the higher the local dielectric constant is, and the better the defects are screened. This phenomena may contribute to the increase of the HMC mobility with gate voltage. This is one possibility, but we do not have direct evidence of this mechanism at play. Another possibility is that there is intermixing and atomic disorder at the interface. This has been heavily studied in \LAO/\STO hetero-structures by TEM or Surface XRD (see for instance Pauli {\em et al.} [S3]). Intermixing has been also evidenced in $\mbox{LaVO}_3$/\STO structures [S4]. Calculations on the \LTO/\STO interface show that La/Sr intermixing on one monolayer is energetically favorable [S5], which can explain optical measurements [S6]. In that case, we expect stronger scattering at the interface (for LMC) than deeper in the substrate (for HMC).\\

\textbf{Part III : Calculation of conduction band profile}\\

Coupled Schr\"odinger and Poisson equations are solved numerically at the interface as proposed in reference [S7], in the effective mass approximation, with quantized electronic sub-bands. The \LTO side of the interface is modelled as an infinite barrier, since the electron gas is most likely on the \STO side. We consider a single parabolic band with a transverse effective mass taken as $m_t=0.7 m_0$ and a confinement mass of $m_z=14 m_0$ which correspond to the $3d_{xy}$ conduction band of \STO [S8].  The confinement mass of the two other bands  ($3d_{xz}$,$3d_{yz}$) would then be $m_z=0.7 m_0$, leading to high energy splitting of the corresponding sub-bands. Hence, they are not considered in the scope of these calculations. We have used here the band calculations value for the light effective mass, which is in agreement with ARPES measurements on \STO surface [S9]. Other measurements on \STO based structures give slightly different values, ranging from 1.2 to 2$m_0$ (see table in [S10]). Choosing one or another value would not strongly affect the main conclusions of the present calculation. Note that ARPES [S9] or ellipsometry [S11] measurements show an extended gas that can be modelled with a rather simple and spatially homogeneous band structure, similar to the one used in this study. The boundary condition of the Maxwell-Gauss equation at the back of the \STO substrate is imposed by the gate voltage. We have therefore a continuous solution for the potential within the whole substrate which is realistic and essential to catch the physics of the system.\\

The electric field dependence of \STO has been modelled as $\epsilon_R(F) = \epsilon_R(F=\infty) + \frac{1}{(A+BF)}$ where $F$ is electric field strength. The temperature dependence of $A$ and $B$ parameters have been measured by Neville \emph{et al.} [S12]. However, since the electron density is quite high, the electrical field at the interface is substantially higher than the measurement range of \STO dielectric constant in the literature, hence the $\epsilon_R(F=\infty)$ parameter is  unknown and usually taken as 1 in most of the calculations. Instead of this unrealistic value for an oxide, we have  used the smallest known value for \STO, which is its high temperature value $\epsilon_R = 300$ (Supplementary figure 3). Furthermore, since the regular Poisson's equation is not suited for spatially varying $\epsilon$, the Maxwell-Gauss equation (which is it's parent equation) was taken with varying $\epsilon$ : $\nabla[ \epsilon (F(z)) F(z)] = \rho(z) $ then derived for both $\epsilon(F)$ and $F$. For  more information on the dielectric constant, see  Supplementary part IV. The last and important ingredient of this calculation is the charge distribution at the discontinuity between the last \TiO plane of \STO (charge 0) and the first \LO plane of \LTO (charge +1), theoretically 6.6$\times 10^{14}\mbox{cm}^{-2}$. However, to account for charge equilibrium, this charge has been set to the measured carrier density at low temperature just after cooling.\\

Using this set of parameters, Schr\"odinger's envelope equation is solved numerically from the input conduction band profile ${E_{\mathrm{C(in)}}}$ to find both energy levels and wave functions. Energy levels are filled to a self consistent Fermi energy $E_{\mathrm{F}}$ to match the sheet electronic density. Maxwell-Gauss equation is then integrated with the computed electronic density profile to give the output conduction band profile ${E_{\mathrm{C(out)}}}$. Both equations are solved iteratively, with the conduction band of the \emph{n}th iteration ${E_{\mathrm{C(in)}}^{(n)}}$ being $(1-f){E_{\mathrm{C(in)}}^{(n-1)}} + f {E_{\mathrm{C(out)}}^{(n-1)}}$, starting from a trial profile ${E_{\mathrm{C(in)}}^{(0)}}$. A deceleration factor $f$ taken as 0.02 ensured smooth convergence generally without oscillations until the sum of the squared error between ${E_{\mathrm{C(in)}}^{(n)}}$ and ${E_{\mathrm{C(out)}}^{(n)}}$ was less than $10^{-9}$~$\mbox{eV}/\mbox{\AA}$.\\
\indent This model is 2D by construction, since we used quantized sub-bands in the z-direction. We checked that, for every gate voltage presented here, the extension of the calculated wave-function of a given energy level is greater than the corresponding Fermi wave-vector. This insures the self consistency of the 2D calculations.
\\

\begin{figure}[h!]
\includegraphics[width=10cm]{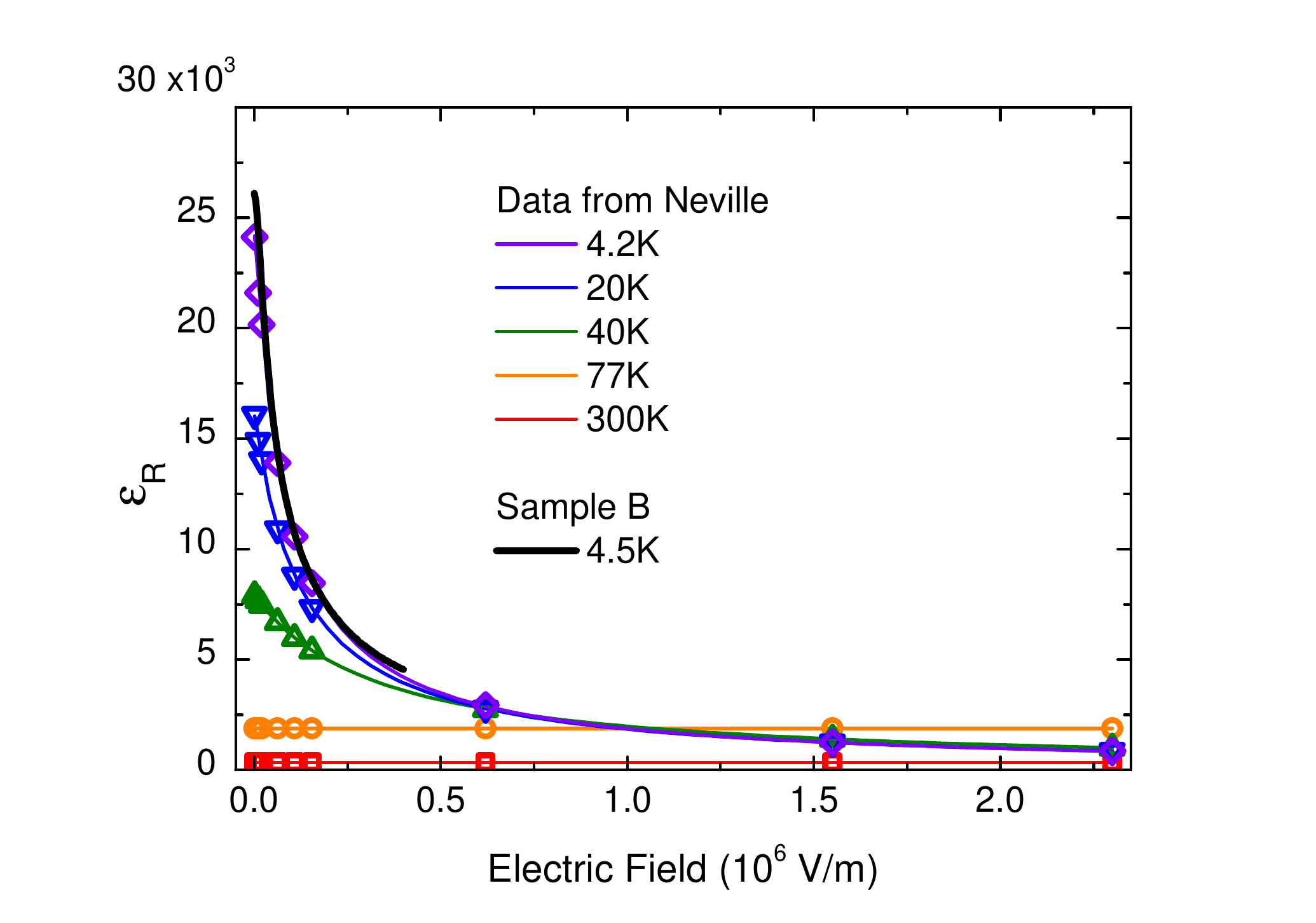}
\caption{ Dielectric properties of \STO. Dielectric constant $\epsilon_R$ of (001) oriented \STO as a function of the electric field for different temperatures. Symbols correspond to data taken from Neville {\em et al.} \cite{NEVILLE:1972p3397} and solid lines to fits according to the formula $\epsilon_R(F) = \epsilon_R(F=\infty) + \frac{1}{(A+BF)}$ (see text). We have added our own measurements of $\epsilon_R$ extracted from the capacitance of sample B at 4.5~K (black line).}
\end{figure}

\textbf{Part IV : Dielectric constant of \STO}\\

As reported in the literature [S12,S14], the \STO dielectric constant $\epsilon_R$ is highly non linear, and depends both on the electric field $F$ and the temperature $T$. Supplementary figure 3 shows $\epsilon_R(F,T)$ measured by Neville \emph{et al.} on (001) oriented \STO, that we used in our calculations. Our own data of $\epsilon_R$ measured on sample B at 4.5~K perfectly matches the data from reference [S12]. According to our model, the electric field in the potential well is in the range where $\epsilon_R$ strongly varies at low temperature. Above 60~K, $\epsilon_R$ is roughly field independent, and rather low : the carriers are confined in the potential well whatever the gate voltage is. This explains the temperature dependence of the Hall carrier density and mobility described in Fig.3 of the main text.\\

Supplementary figure 4a,b,c shows the electric field and the dielectric constant calculated in the band bending model presented in the main text (Fig.4a,b,c). As the carrier densities are quite high, the electrical field at the interface is far beyond the measurement range presented in supplementary figure 3, which does not allow the determination of the $\epsilon_R$ value at such high fields. A realistic value of $\epsilon_R=300$, which is it's smallest known value, has been taken in the calculations, as argued in supplementary part III. The panel \textbf{d} of Supplementary figure 4 is a blow-up of the inset of the Fig.4.a presented in the main text. It displays the conduction band profile for negative, null and positive back-gate voltage $V_{\mathrm{G}}$. In the later case, a maximum of the conduction band bending occurs, which limits the density of negative carriers that the well can accommodate.\\

\begin{figure}[tb]
\includegraphics[width=10cm]{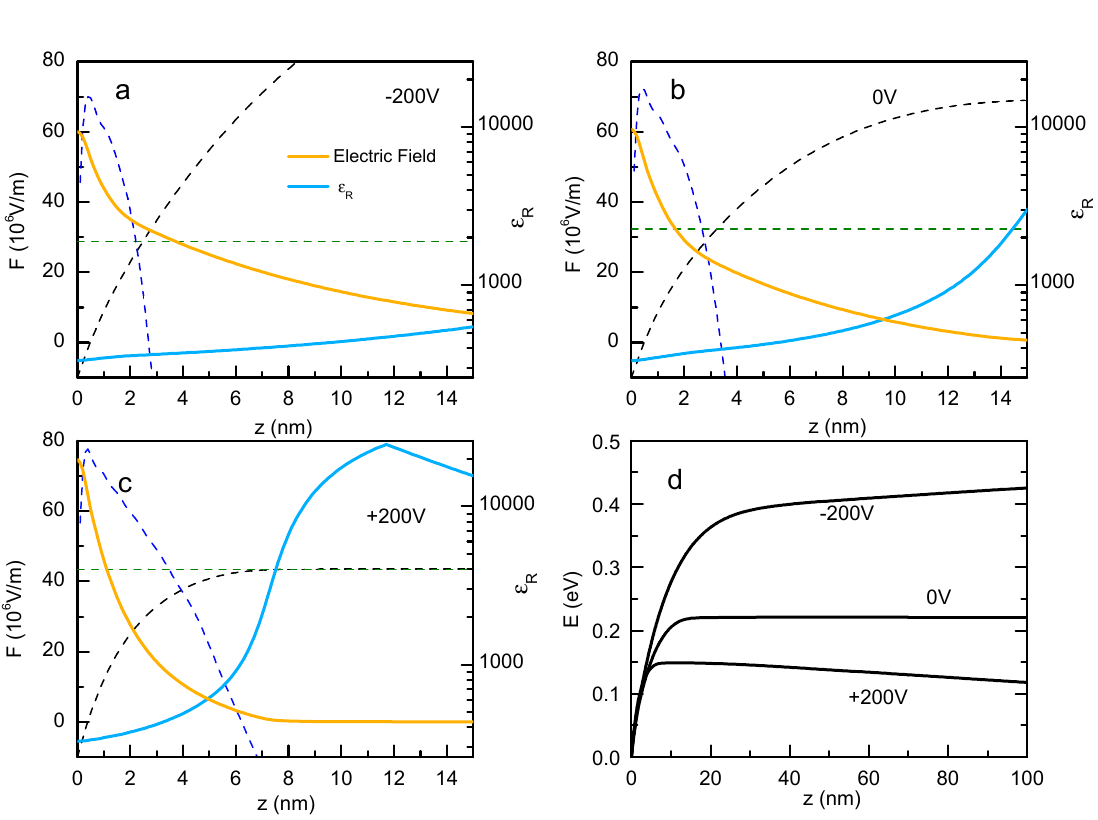}
\caption{ \large {\bf a}, {\bf b} and {\bf c}, Additional information for band bending calculations presented in the main text fig.4. Figures show the electric field along the z-axis perpendicular to the interface (left scale) and the dielectric constant $\epsilon_R$ (right scale) as a function of depth from the interface. The conduction band $E_{\mathrm{C}}$ (dashed black line), the Fermi level $E_{\mathrm{F}}$ (dashed green line) and the carrier density $n$ (dashed dark blue line) are also shown as in Fig.4 of the main text (see main text for scales). {\bf d}, Conduction band profile as a function of depth from the interface on an extended scale, for the gate voltages presented in panels {\bf a}, {\bf b} and {\bf c}. }
\end{figure}

In our calculation, the potential is rapidly varying over short distances in the well. Therefore, in principle, in addition to the non-linear electric field dependence, it would be suitable to also introduce a $q$-dependence in the expression of $\epsilon_R(F,q)$. This quantity has not been directly measured yet. At zero field, the $q$-variation of $\epsilon_R$ can be deduced from the dispersion of the soft transverse optical phonon related to polarisation fluctuations [S13]. A strong $q$-dependence of the phonon energy is observed for $q$ in the range 1 to 2 nm$^{-1}$ mainly. In our calculation, the width of the quantum well corresponds to characteristic wave vectors of the order of 0.2 nm$^{-1}$. According to the figure 2 of reference [S13], the phonon energy at this wave-vector is close to the $q$=0 one. Moreover, the spatial variation of  $\epsilon_R$ in the well are expected to be strongly reduced by the electric field (supplementary figure 4). Therefore, using the $\epsilon_R(F,q=0)$ value is a reasonable approximation in this context.\\

\begin{figure}[h!]
\includegraphics[width=10cm]{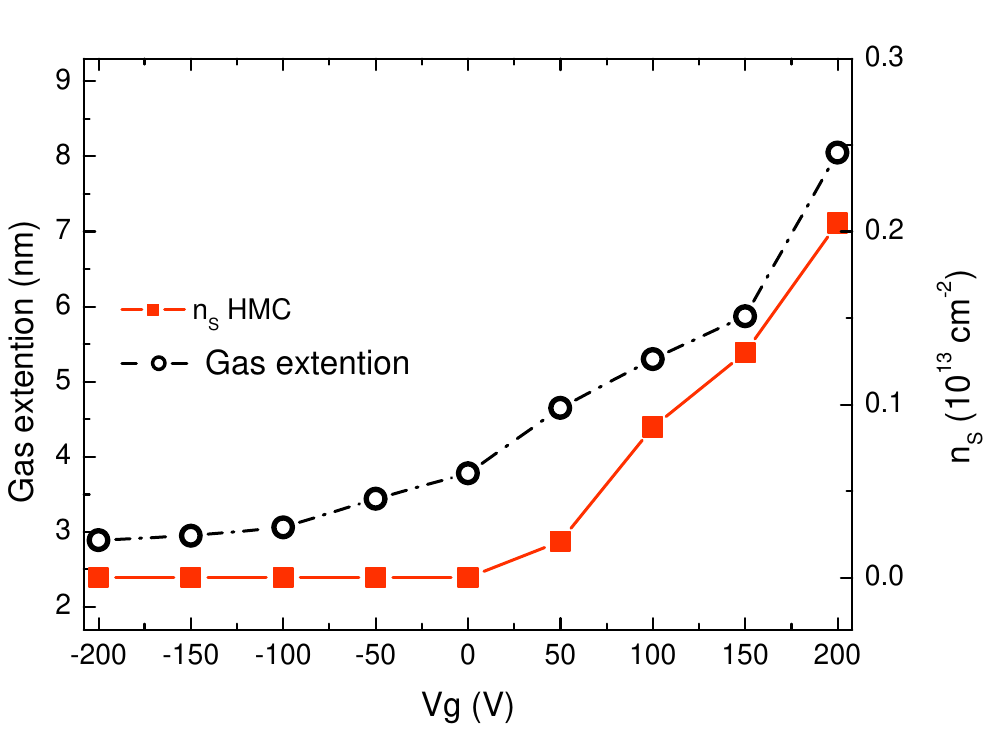}
\caption{Gas extension as a function of the gate voltage according to our model for sample B. The right scale displays the Highly Mobile Carrier (HMC) density given by high field Hall measurements and two carriers analysis, as a function of the gate voltage for sample B. HMC density rises when the gas extension increases.}
\end{figure}

\textbf{Part V : Band bending calculation on sample B}\\

We present here the result of band bending calculations for sample B shown in Fig.3 of the main text. When cooling down, its carrier density is slightly higher than sample A. This initial difference obviously persists with gate voltage.  For example, the carrier density for sample B at $V_{\mathrm{G}}=$-200~V is equivalent to that of sample A at $V_{\mathrm{G}}=$+50~V. However, as shown on supplementary figure 5, its behaviour is rather similar, and HMC start appearing for positive voltage only. Our self-consistent calculation shows that this also corresponds to the extension of the gas on the shallow side of the well. \\

\textbf{References of the Supplementary Material section.}\\

\noindent \lbrack S1\rbrack  D. S. Deak et al., J. Phys. Chem. B,  { \bf 110} 9246, (2006)\\ 
\lbrack S2\rbrack S. S. A. Seo et al., Appl. Phys. Lett. { \bf 95}, 082107 (2009).\\
\lbrack S3\rbrack S. A. Pauli et al.,Phys. Rev. Lett. {\bf 106}, 036101 (2011).\\
\lbrack S4\rbrack  L. Fitting Kourkoutis et al. Appl. Phys. Lett. {\bf 91}, 163101 (2007).\\
\lbrack S5\rbrack  J. J. Pulikkotil et al., Appl. Phys. Lett. {\bf 99}, 081915 (2011).\\
\lbrack S6\rbrack  S. S. A. Seo et al., Phys. Rev. Lett. { \bf 99}, 266801 (2007).\\
\lbrack S7\rbrack  F. Stern, Phys. Rev. B {\bf 5}, 4891--4899 (1972).\\
\lbrack S8\rbrack  L. F. Mattheis, Phys. Rev. B { \bf 6}, 4718--4740 (1972).\\
\lbrack S9\rbrack  W. Meevasana et al., Nature Mater. {\bf 10}, 114--118(2011).\\
\lbrack S10\rbrack M. Kim, C. Bell, Y Kozuka, M. Kurita, Y. Hikita, H. Y. Hwang, Phys.Rev. Lett.  {\bf 107}, 106801 (2011).\\
\lbrack S11\rbrack A. Dubroka et al., Phys. Rev. Lett.{ \bf 104}, 156807 (2010).\\
\lbrack S12\rbrack R. Neville et al., J. Appl. Phys.{ \bf 43}, 2124--2131  (1972).\\
 \lbrack S13\rbrack Y. Yamada and G. Shirane, J. of the Phys. Soc. of Japan { \bf 26}, 396 (1969).\\
\lbrack S14\rbrack J. Hemberger et al., Phys. Rev. B{ \bf 52}(18), 13159--13162 (1995).\\


\begin{thebibliography}{1}
\bibitem{noteAniso20} At negative voltage, the anisotropy of the resistance is stronger (see suplementary part II), showing that the scattering is more anisotropic. This is an indication that the 2DEG is close to the surface where surface reconstructions occurs.
\end{thebibliography}
\end{document}